\documentclass[a4paper,aps,pra,twocolumn,superscriptaddress,showpacs]{revtex4}

\usepackage{graphicx}
\begin{document}

% \draft command makes pacs numbers print
%\draft
% repeat the \author\address pair as needed
\title{Precessing vortices and antivortices in ferromagnetic elements}
\author{A. Lyberatos}
\affiliation{Department of Materials Science,University of Crete,PO BOX 2208,71003 Heraklion,Greece}
\author{S. Komineas}
\affiliation{Department of Applied Mathematics, University of Crete, 71409 Heraklion, Crete, Greece}
\author{N. Papanicolaou}
\affiliation{Department of Physics, University of Crete, 71003 Heraklion, Crete, Greece}
\affiliation{Institute for Theoretical and Computational Physics, University of Crete, Heraklion, Greece}
%\date{\today}

\begin{abstract}

A micromagnetic numerical study of the precessional motion of the vortex and antivortex states in
soft ferromagnetic circular nanodots is presented using Landau-Lifshitz-Gilbert dynamics. For sufficiently small
dot thickness and diameter, the vortex state is metastable and spirals toward the center of the dot when its initial displacement is smaller than a critical value. Otherwise, the vortex spirals away from the center and eventually exits the dot which remains in a state of in-plane magnetization (ground state). In contrast, the antivortex is always unstable and performs damped precession resulting in annihilation 
at the dot circumference. The vortex and antivortex 
frequencies of precession are compared with the response expected on the basis of Thiele's theory of collective coordinates. We also calculate the vortex restoring force
with an explicit account of the magnetostatic and exchange interaction on the basis of the 'rigid' vortex and 'two-vortices side charges free' models and show that neither model explains the vortex translation mode eigenfrequency 
for nanodots of sufficiently small size.  
\end{abstract}

% insert suggested PACS numbers in braces on next line

%Keywords: Magnetic vortex, antivortex,  magnetic dots, vortex dynamics. 


%\pacs{heat-assisted magnetic recording, perpendicular media, thermal stability }

\maketitle

% body of paper here
\section{Introduction}

The vortex state is one of the equilibrium states of thin soft ferromagnetic elements of micrometer size and below (magnetic dots). The interplay between the magnetostatic and exchange energy favours an in-plane, closed flux domain structure with a $10-20$ nm central core, where the magnetization turns out of plane to avoid the high energetic cost of anti-aligned moments. Core reversal can be triggered by application of an in-plane  pulsed field or pulsed current allowing   
the possibility of application of patterned thin film elements in data storage and magnetic and magneto-electronic random access memory \cite{wae:06}. Core reversal is usually assumed to arise  from the spontaneous  creation of a vortex-antivortex (VA) pair (vortex dipole) of opposite polarity with respect to the original vortex, followed by collision of the pair with the original  vortex. A fundamental understanding of the dynamics of vortices and antivortices is therefore necessary
to control the switching of the magnetization.

 The basic excitation mode of the vortex or antivortex state from its equilibrium position is in-plane gyrotropic motion. 
It is a low frequency  (GHz) mode corresponding to the displacement of the whole structure. The generalized dynamic force  can be determined using
Thiele's collective-variable approach \cite{thi:73}-\cite{hub:82}. Theoretical \cite{pap:91}-\cite{kom:96} and experimental \cite{cho:04} studies of the dynamics of magnetic vortices
in 2D films have shown a connection with the
topology of the magnetization structure . Magnetic vortices confined in circular dots can be  described by analytical models based on different methods for accounting for the magnetostatic interaction \cite{gus:01}-\cite{gus:02}.  The vortex may be 'rigid' or deform  so that no magnetic charges appear at the side of the cylinder.  The latter (two-vortices side charge free model) provides a good description of the dynamic behavior of vortices in submicron-sized permalloy dots, in particular the increase of vortex eigenfrequency with dot aspect ratio $L/R$, where $L$ is the dot thickness and $R$ is the dot radius. The basic assumption in these calculations is that the vortex displacement $l$ from
equilibrium, at the dot center, is small $l<<R$.

The main objective here is to consider the response of the vortex state to large perturbations, induced
 for instance by thermal activation. We focus our attention to dots of sufficiently small radius and thickness so that the vortex state is metastable . Using micromagnetic calculations it is  shown that vortex stability can be defined in terms of a critical displacement $l_{c}$ leading to an irreversible transition
to the ground state characterized by in-plane magnetization. We test the accuracy of the collective coordinate representation for vortex and antivortex dynamics and discuss the limitations of the approximate analytical models
\cite{met:02},\cite{gus:02}.

\section{Vortex and antivortex precession}

The dynamical behavior of a single vortex or antivortex trapped in a ferromagnetic nanodot 
was studied using a finite-difference Landau-Lifshitz-Gilbert micromagnetic model.
The material parameters for permalloy (Ni$_{80}$Fe$_{20}$) were used in the calculations: 
saturation magnetization  $M_{s}=800$
emu/cc and exchange stiffness coefficient $A=1.3 \times 10^{-6}$ erg/cm. The magnetic anisotropy was neglected and the exchange length is $l_{ex}=\sqrt{ A/( 2 \pi M_{s}^{2}) } =5.7$ nm.  The disk thickness is in the range 
$L=5-10$ nm , comparable to the exchange length, so the magnetization dependence along the dot normal can be neglected to a first approximation. The disk radius is not large compared to the vortex core and is initially taken as $R=30$ nm. Inspection of the phase diagram of magnetic ground states for this material \cite{sch:03} indicates that for the particular choice of disk dimensions, in-plane magnetization is the ground state of the system.  The integration of the LLG equation is carried out using a damping parameter $\alpha=0.01$, which is appropriate for permalloy \cite{ber:08}.

The initial vortex or antivortex structure is defined as follows: Using the coordinate system with the z axis parallel
to the dot cylindrical axis, the magnetization components are $m_{z}=\lambda \cos\Theta(\rho), m_{x}+im_{y}=\sin\Theta\exp(i\kappa(\phi-\phi_{o}))$, where ${\bf m(r)}={\bf M(r)}/M_{s}$ is the reduced magnetization, $\rho,\phi$ are cylindrical coordinates, $\lambda=\pm 1$ is the vortex polarity,  
 and $\Theta(\rho)$ is the magnetization angle to the normal (z) direction. The vortex number is $\kappa=1$ for a vortex and 
$\kappa=-1$ for an antivortex structure. The constant $\phi_{o}=\pm \pi /2$ defines the chirality i.e. the direction of the curl of the vortex. The specific choice of chirality is largely insignificant for vortex
dynamics, in contrast to the vortex number $\kappa$ and polarity $\lambda$ which play an important role.

\begin{figure}[t]
   \centering
   \includegraphics[width=2.7in]{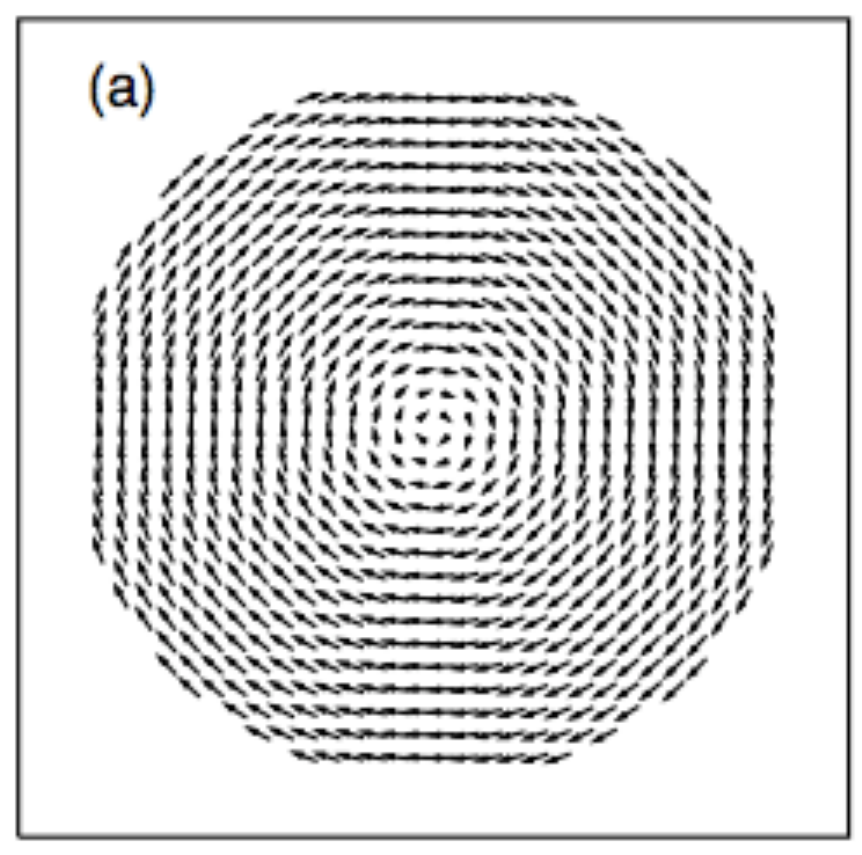}
   \includegraphics[width=2.7in]{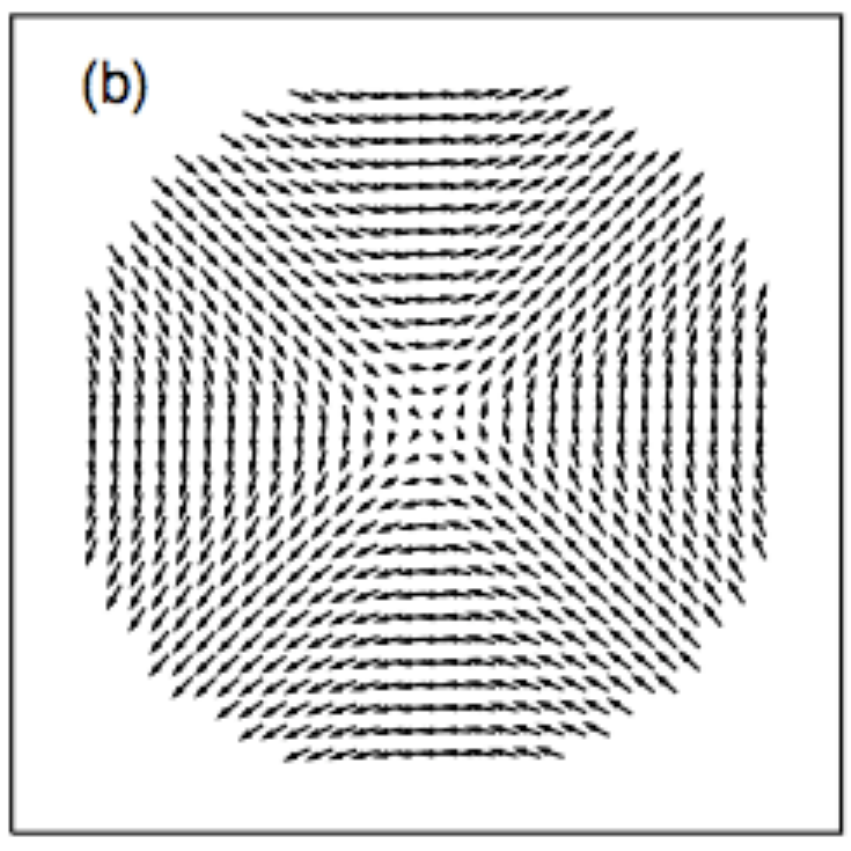}
\caption{a) Vortex snd b) antivortex structure in permalloy dot of thickness $L=5$ nm and
radius $R=30$ nm.  }
   \label{fig:fig1}
\end{figure}

The initial profile $\Theta (\rho)$ is assumed to have the model form $\cos\Theta=(\cosh (c\rho/l_{ex}))^{-1}$ which is appropriate for easy-plane ferromagnets  \cite{kom:09}. The parameter $c$ defines the core radius
$b$. The core radius is determined by the dot thickness $L$ and the exchange length $l_{ex}$ but is 
independent of the dot radius \cite{shi:00}. In the following, dot thickness is $L=5$ nm, unless  specified otherwise. Numerical calculations of relaxed vortex and antivortex structures at the dot center were found to be in good agreement
with the model profile for $c  \simeq 1$.
 Typical examples of a vortex and antivortex structure are shown in  
Fig.~\ref{fig:fig1}. It should be noted that other choices for the vortex profile are possible \cite{lan:05}, for instance
  $\tan(\Theta /2) =\rho /b$ was obtained by Usov {\em et al.} \cite{uso:93} using a variational
procedure to minimize the exchange energy whereas the vortex core radius $b$ was determined by minimization also of the magnetostatic energy. 
This form is employed in analytical calculations of vortex states in ferromagnetic disks \cite{gus:01}-\cite{gus:02},
 however, micromagnetic modelling by Scholz {\em et al.} \cite{sch:03}, confirmed by our calculations,
has shown that it underestimates 
the core radius $b$, defined in Ref.\cite{sch:03} as the first $m_{z}=0$ crossover from the vortex center.   

\begin{figure}[t]
  \centering
   \includegraphics[width=3.5in]{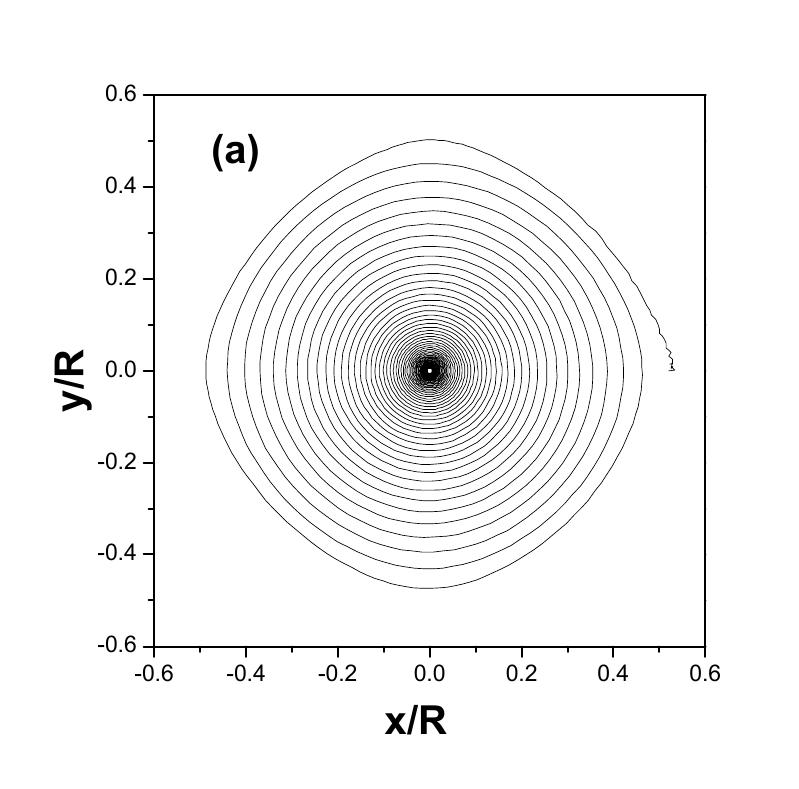}
   \includegraphics[width=3.5in]{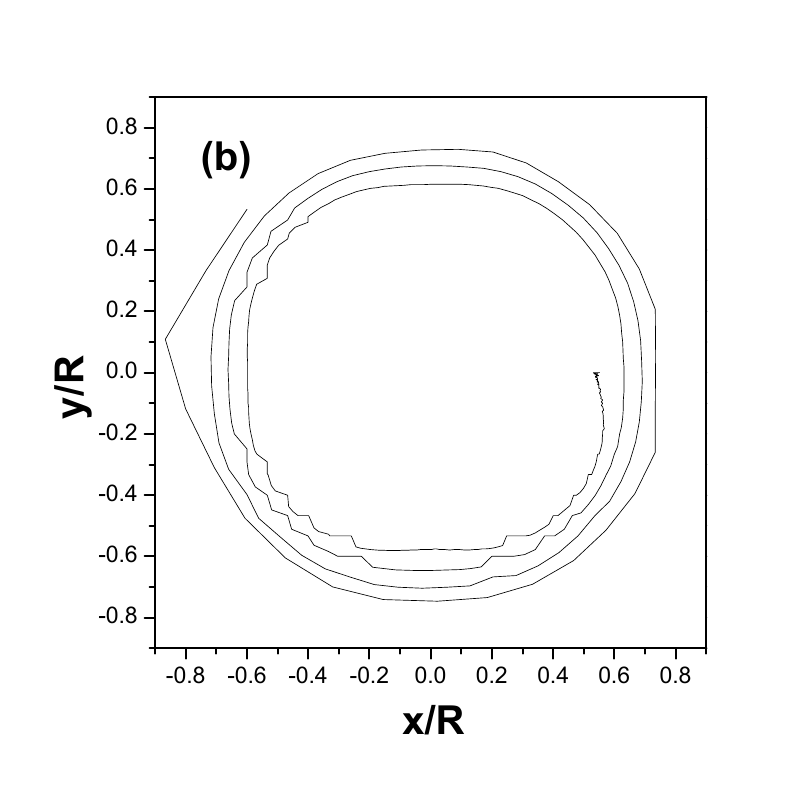}
\caption{Trajectory of a vortex of positive polarity $\lambda=1$ in zero field for a time interval $0<t<2\times 10^{4}$.
The initial position vector of the vortex is a) $l_{o}=(0.52 R,0)$ and b) $(0.53 R,0)$. The anisotropy is neglected and the damping is $\alpha=0.01$.   }
   \label{fig:fig2}
\end{figure}

The dynamical behavior of the magnetic vortex $(\kappa =1 )$  depends on the initial displacement $l(t=0)=l_{o}$ from the disk center.
If the displacement is equal or smaller than some critical value $l_{c}=0.52 R$, the damped  precession of the vortex leads to relaxation to the disk center.  For positive vortex polarity $\lambda=1$, an anticlockwise precession is observed
 as shown in Fig.~\ref{fig:fig2}a. At the start of the simulation the precession is not smooth as a result of the internal relaxation 
(deformation) of the model vortex to minimize the energy during precession.  If $l_{o} > l_{c}$, the damped precession is clockwise and the distance from the dot center
increases, as shown in Fig.~\ref{fig:fig2}b, until the vortex is annihilated and the magnetization is aligned along the in-plane direction with quasi-uniform magnetization, the so-called 'leaf' state \cite{met:04}  which is the ground state of the system. Irregularities in the precessional motion arise from the uncertainty on the position of the vortex. 

The antivortex instead is always unstable. For any choice of initial displacement $l_{o}$, the antivortex performs damped precession  to the edge of the disk  and is annililated. For positive polarity, anticlockwise precession is observed. 
The sense of gyrotropic motion of a vortex or antivortex is switched on reversal of the polarity.

\section{The collective coordinate approach}

The damped precession of the vortex or antivortex can be described using  
Thiele's equation \cite{thi:73},\cite{hub:82} augmented by a dissipative term.

\begin{equation}
{\bf G} \times \frac{d{\bf l}}{dt}+2QG \eta \frac{d{\bf l}}{dt}-\frac{ \partial E({\bf l})} {\partial {\bf l}} =0
\label{equ:thiele}
\end{equation}
where ${\bf l}=(l_{x},l_{y})$ is the  position of the vortex center and $E(l_{x},l_{y})$ 
is the potential energy of the shifted vortex. 
The gyroforce ${\bf G} \times d{\bf l}/dt$  depends on the topological structure of the magnetization and is proportional to the gyrovector ${\bf G}=-G \hat{{\bf z}} $, where the gyroconstant
is $G=2\pi \kappa \lambda L M_{s}/\gamma$ and $\gamma=1.76 \times 10^{7} $ rad Oe$^{-1}$ s$^{-1}$  is the gyromagnetic ratio.  $Q=-\frac{1}{2}\kappa\lambda$ is the skyrmion number and $\eta$ is the dissipation constant.

 For axially symmetric energy potential $E=E(l)$ where $l=\sqrt{l_{x}^{2}+l_{y}^{2}}$ , $\partial E/\partial l_{x}=E' l_{x}/l$ and it is straightforward to show that

\begin{equation}
\dot{l}_{x} -2Q \eta \dot{l}_{y}  =   -\omega l_{y} 
\end{equation}
\begin{equation}
2Q\eta \dot{l_{x}}+ \dot{l_{y}}  =    \omega l_{x} 
\end{equation}
where the angular frequency is 
\begin{equation}
\omega =\frac{1}{Gl} \frac{\partial E}{\partial l}
\label{equ:eg}
\end{equation}
 
The vortex motion in complex form is given by

\begin{equation}
\left( 1+2Q\eta i \right) \left( \dot{l_{x}} +i \dot{l_{y}} \right) =i\omega \left( l_{x} +i l_{y} \right)
\end{equation}

Introducing polar coordinates $l_{x}+i l_{y}=l e^{i\phi}$ 

\begin{equation}
\dot{l}=\frac{2Q\eta}{1+\eta^{2}} \omega (l) l
\end{equation}
\begin{equation}
\dot{\phi}=\frac{\omega(l)}{1+\eta^{2}}
\label{equ:om}
\end{equation}

Dividing and integrating over the time interval of the damped precession, the time dependence  of the
precession angle is 

\begin{equation}
\phi(t)=\frac{2Q}{\eta} \ln \left( \frac{l}{l_{o}} \right) 
\label{equ:phi}
\end{equation}
where $l_{o}$ is the initial displacement of the vortex center.
The clockwise or anticlockwise sense of gyration is therefore dependent on the skyrmion number $Q$. 

\begin{figure}[t]
  \centering
   \includegraphics[width=3.5in]{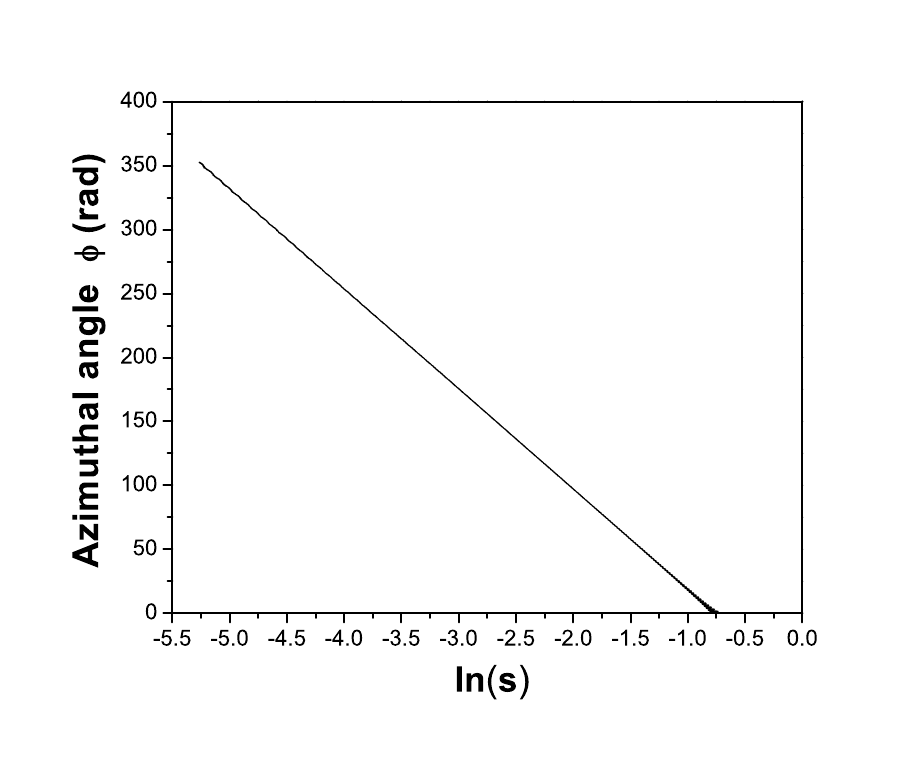}
\caption{Azimuthal angle $\phi$ of the vortex position as a function of the logarithm of the reduced vortex
displacement $s=l/R$. The initial radial position is $l_{o}=0.52 R$. }
\label{hamrtrackfig3}
   \label{fig:fig3}
\end{figure}

Micromagnetic simulations of vortex motion were performed and the position of the vortex center $(l_{x},l_{y})$
was determined by a method of interpolation for the position of maximum
$m_{z}$. The precession angle 
$\phi=\arctan(l_{y}/l_{x})$ was found to vary linearly with the logarithm of the vortex shift, in agreement
with Eq.(\ref{equ:phi}). Fig.~\ref{fig:fig3} shows numerical data for damped precession of a vortex of positive polarity $(Q=-1/2)$
 with initial displacement $l_{o}=0.52 R$. The relaxation to the disk center involves many revolutions (Fig.~\ref{fig:fig2}a) and
the gradient  $-1/\eta $ provides an accurate estimate of the dissipation constant $\eta=0.013$.  It is evident that for permalloy nanodots, the damping in the vortex motion 
is weak and the angular frequency of precession in Eq.(\ref{equ:om}) can be approximated 
using $\omega \simeq d\phi/dt$.
 
 \begin{figure}[t]
  \centering
   \includegraphics[width=3.5in]{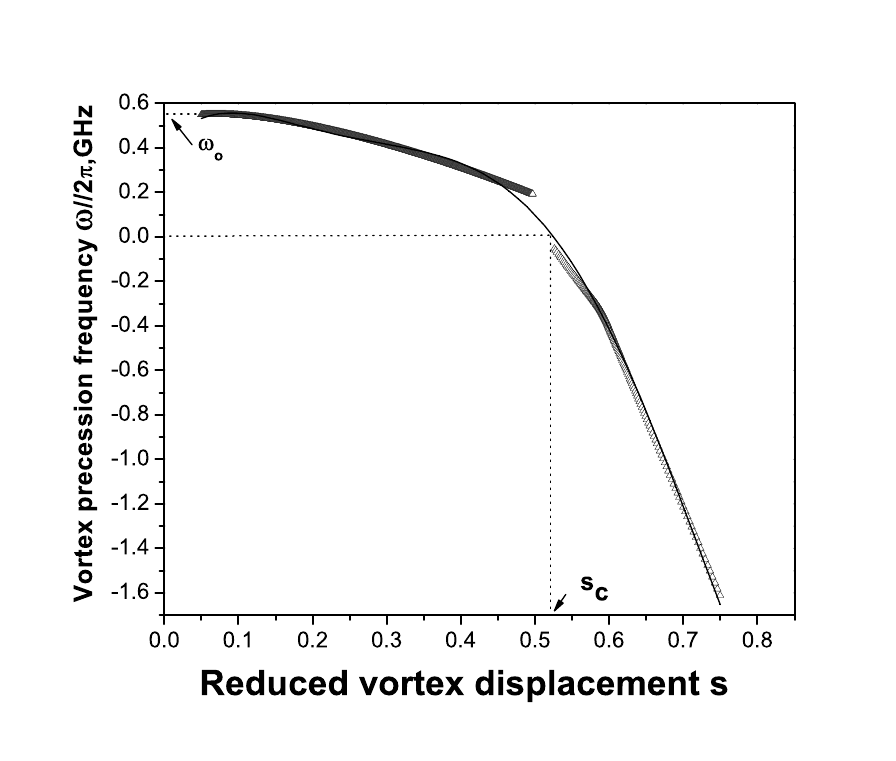}
\caption{Precession frequency of a vortex $\omega/2\pi$ as a function of the reduced diplacement $s=l/R$
from the center of the dot.  }
   \label{fig:fig4}
\end{figure}

Fitting the numerical $\phi(t)$ curve to a 4th order
polynomial, the time variation of the angular frequency $\omega(t)$ can be determined.
 The vortex shift $l(t)$ exhibits oscillations that are neglected by fitting to a 4th order polynomial and the  $\omega(l)$ dependence is obtained using the $\omega(t)$ curve. The same procedure is employed for damped precession leading to vortex annihilation ($l_{o}=0.53 R$). The combined results for the dependence of the frequency of precession 
$\omega/2\pi$ on reduced vortex shift $s=l/R$ are illustrated in Fig.~\ref{fig:fig4}. 

For small displacement of the vortex center from its equilibrium position $(l=0)$, the potential energy 
is $E(l)=E(0)+(1/2) \kappa l^{2}$, where $\kappa$ is the stiffness coefficient  and the eigenfrequency is  
$\omega_{o}=\kappa /G$ \cite{gus:02}. At the critical displacement $l_{c}$, corresponding to a maximum in the
potential energy $E(l)$ the precession frequency vanishes.

The motion of vortices and antivortices is driven by the restoring force $\partial E/\partial {\bf l}$ 
(Eq.\ref{equ:thiele}). The potential energy of the shifted vortex is axially symmetric $E=E(l)$ and can be written
\begin{equation}
E(l)=E(0)+G\int_{0}^{l}\omega(\rho) \rho d\rho
\end{equation}

\begin{figure}[t]
  \centering
   \includegraphics[width=3.5in]{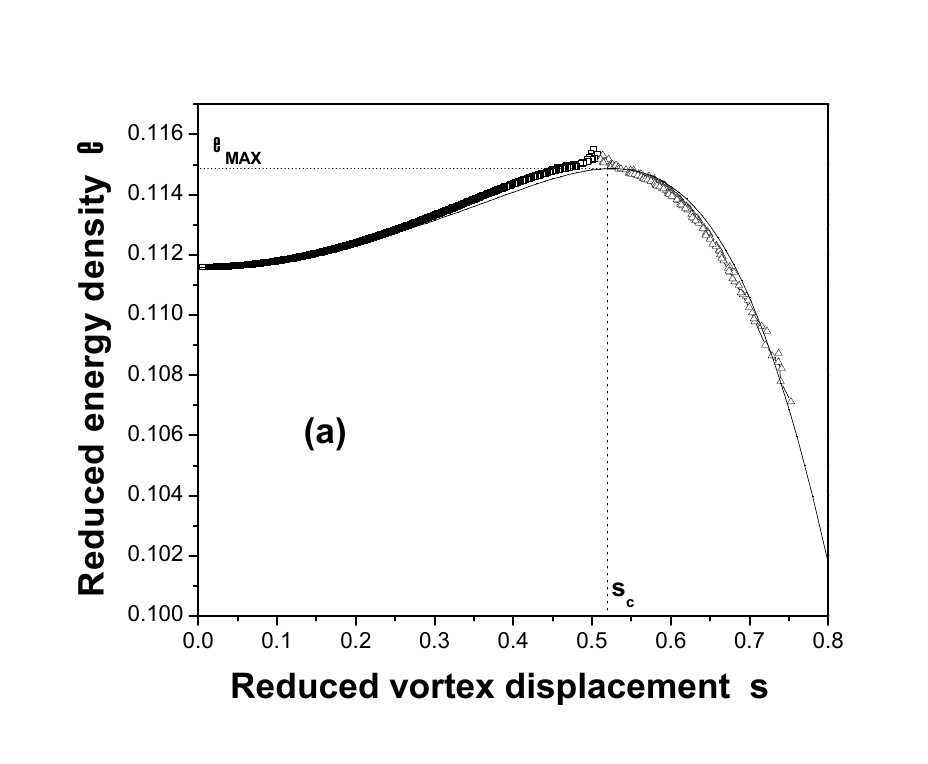}
   \includegraphics[width=3.5in]{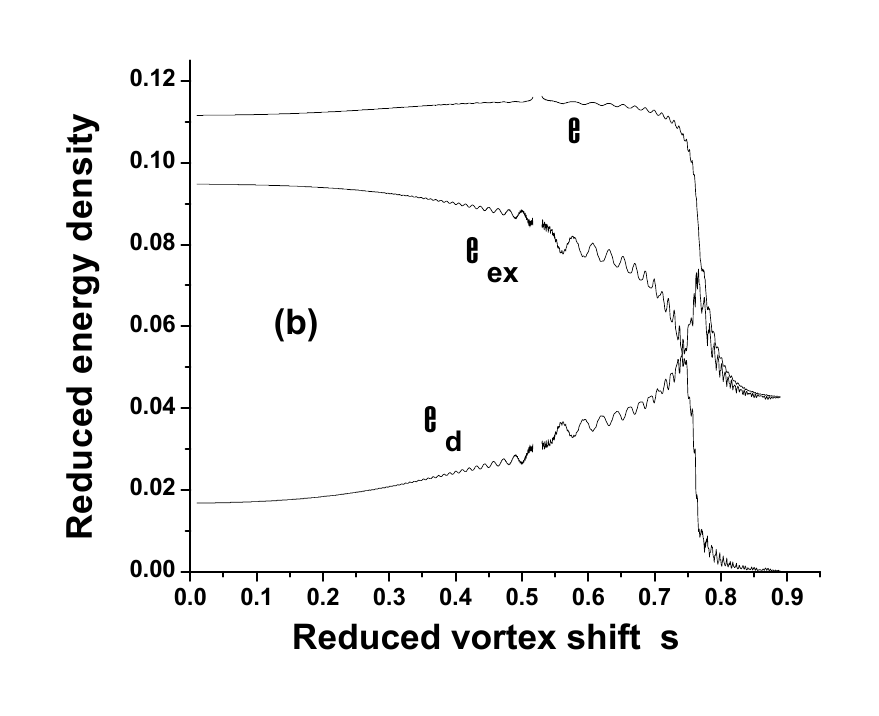}
\caption{(a) Reduced energy density of a permalloy dot $\epsilon=E/4\pi M_{s}^{2}V$ as a function of normalized
 vortex displacement $s$. Results are shown for two sets of micromagnetic simulations (markers) 
and analytical (solid line) calculations using Thiele's collective variable theory (Eq.\ref{equ:ee}).
The maximum energy density $\epsilon_{max}$ occurs at vortex displacement $s_{c}$.
(b) The contribution of the magnetostatic and exchange terms to the total energy, obtained
from micromagnetic calculations.
 }
   \label{fig:fig5}
\end{figure}

A simpler form in terms of the reduced energy density  $\epsilon=E/4\pi M_{s}^{2}V$ over the dot volume $V$ is
\begin{equation}
\epsilon(l)=\epsilon(0)-4 Q \int_{0}^{s}  \overline{\omega}  (s ') s ' 
ds '
\label{equ:ee}
\end{equation}
where $s '=\rho/R$, $s=l/R$ is the reduced   
displacement of the vortex and the time associated with $\overline{\omega }$ is  scaled by $\tau_{o}=1/4\pi\gamma M_{s}$. 
The function $\overline{\omega}(s ')$ was determined from micromagnetic simulations of the vortex precession and using Eq(\ref{equ:ee}) the curve $\epsilon(l)$ expected from application of the theory of collective coordinates,
was obtained, as shown in Fig.~\ref{fig:fig5}a. Superimposed is the energy evaluated directly from simulations of vortex relaxation. 
Thiele's theory appears to provide a good description of the vortex precession despite its limitations, for instance,
 a) Thiele's approach is known to be an approximate description of vortex dynamics in infinite films b) the vortex is here confined in a nanodot ($R=5.3 l_{ex}$) and c) the vortex structure does not remain rigid during the relaxation process but is modified as
a result of the change in the distribution of the demagnetizing fields. The magnetostatic and exchange contribution to the energy variation $\epsilon(l)$ was obtained from micromagnetic calculations and is shown in Fig.~\ref{fig:fig5}b. 
It should be noted that incorporating the demagnetizing energy to the total anisotropy, is strictly valid for infinite
thin films and results in  monotonically decreasing energy $\epsilon(l)$, as reported in Ref. \cite{she:05}. 
The oscillations in the energy variation during precession are related to the finite micromagnetic grid.

The potential energy of the vortex attains a maximum value at some critical value of the displacement $s_{c}=0.52$ 
corresponding to a zero crossover of the precession frequency $\omega$ (Fig.~\ref{fig:fig4}). The stability of the vortex at the dot center arises from the magnetostatic energy, in particular the volume magnetic charges resulting from vortex deformation 
 and the surface charges at the side of the cylinder \cite{gus:02}. The face charges  do not depend on $s$ since the charge distribution on the top and bottom surfaces of the disk is unchanged with the vortex displacement. The exchange energy
decreases with increasing vortex shift $s$ \cite{gus:02}. The  magnetostatic and exchange
contributions to the restoring force are in exact balance at the point of maximum energy. 

\begin{figure}[t]
  \centering
   \includegraphics[width=3.5in]{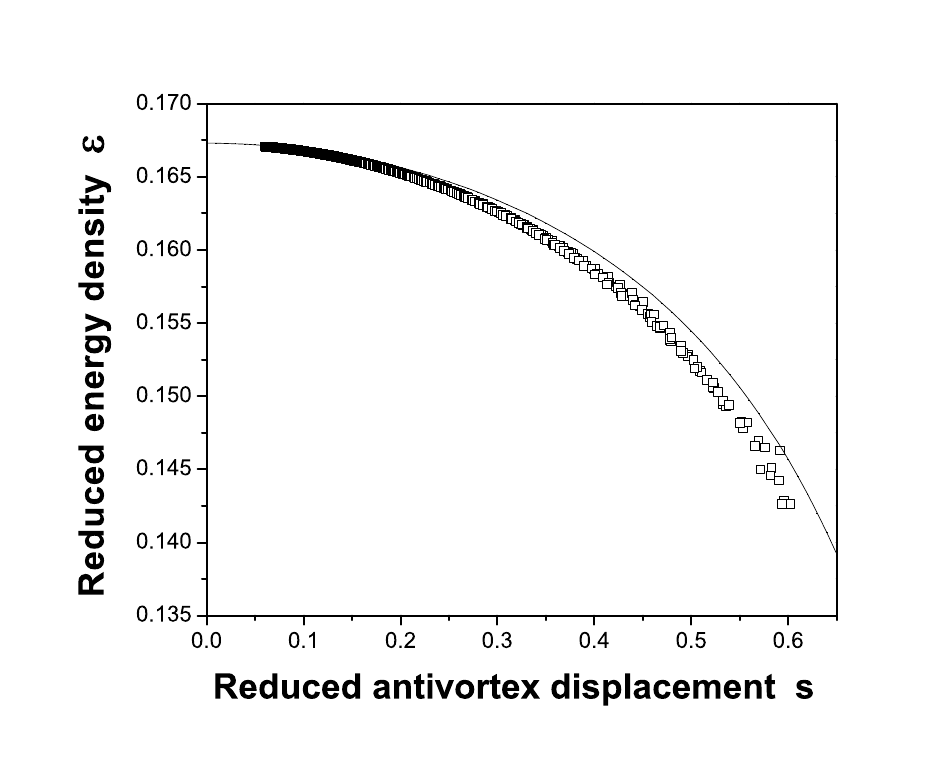}
\caption{Energy density  $\epsilon=E/4\pi M_{s}^{2}V$  vs antivortex displacement 
 $s$. The notation is similar to Fig.~\ref{fig:fig5}a. }
   \label{fig:fig6}
\end{figure}

\begin{figure}[t]
  \centering
   \includegraphics[width=3.5in]{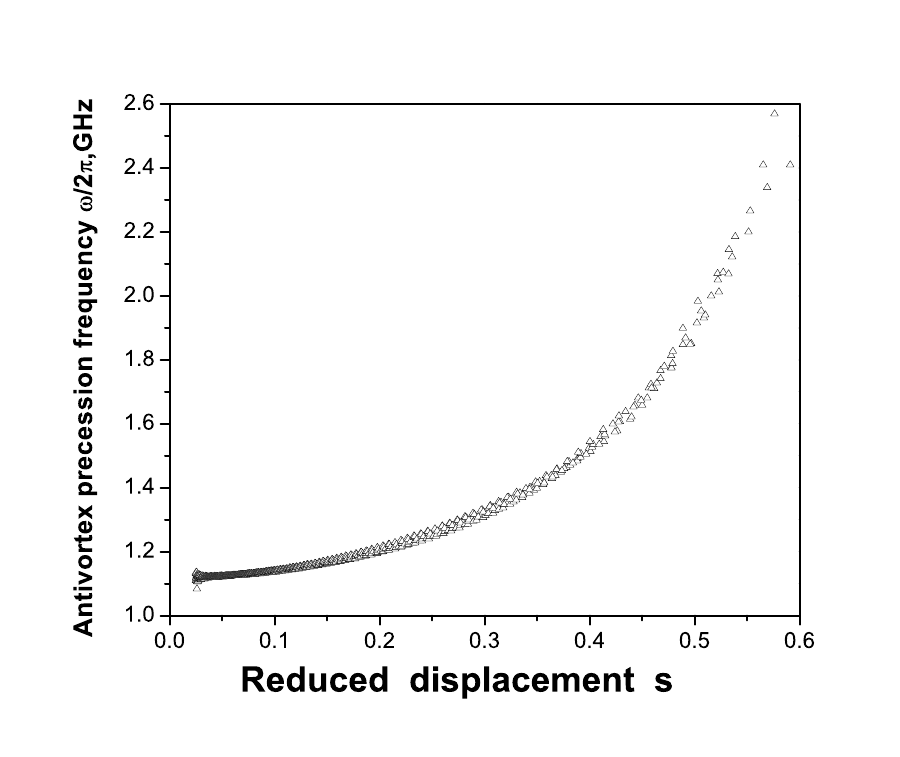}
\caption{Antivortex precession frequency $\omega/2\pi$ as a function of the reduced diplacement $s=l/R$
from unstable equilibrium position at the dot center.}
   \label{fig:fig7}
\end{figure}

A similar analysis was carried out for an antivortex structure in a dot of identical dimensions. The potential energy decreases monotonically with increasing displacement s, as shown in Fig.~\ref{fig:fig6}. Application of the collective coordinates
treatment results in the solid curve in Fig.~\ref{fig:fig6} of slightly smaller curvature.
The vortex instability arises from the uncompensated magnetic charge distribution within the antivortex core (Fig.~\ref{fig:fig1}b),
so the magnetostatic energy is reduced upon motion away from the dot center.
  The precession frequency increases during the relaxation process as shown in Fig.~\ref{fig:fig7}, as a result of the steeper energy
gradient for large displacement $s$. Assuming identical position, it is evident that the antivortex precesses faster than
the vortex as a result of the larger magnetostatic energy gradient.  

\section{Dependence of vortex precession on disk size}

\begin{figure}[t]
  \centering
   \includegraphics[width=3.5in]{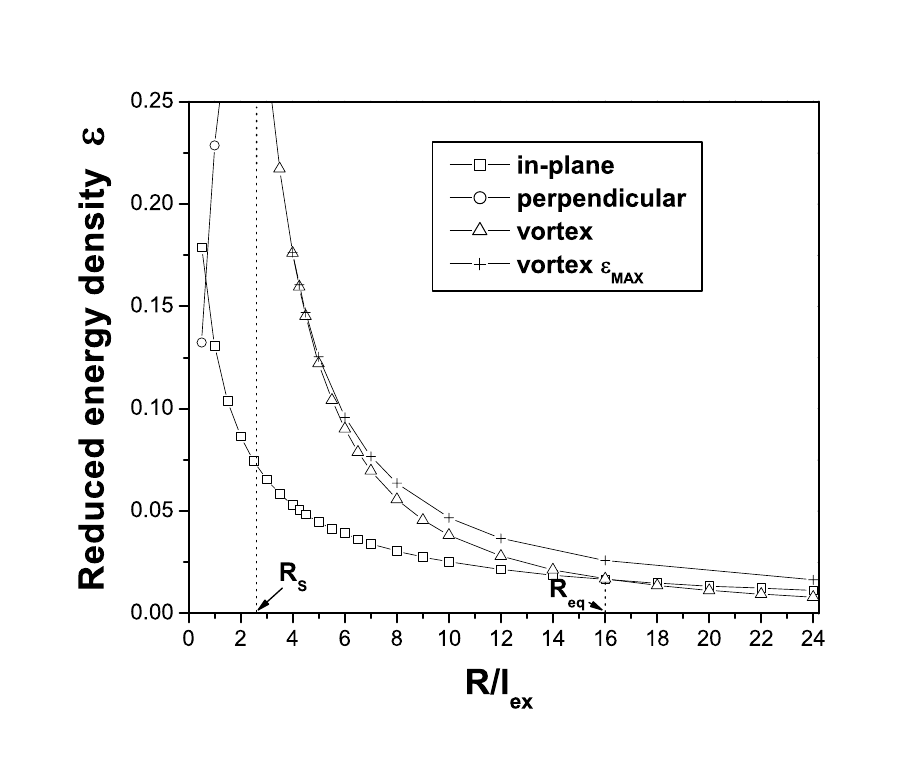}
\caption{Micromagnetic calculations of scaled dot energy density $\epsilon$ 
vs dot radius $R$ (in units of $l_{ex}$) for the three equilibrium
states of the magnetization and the vortex state of maximum energy. }
   \label{fig:fig8}
\end{figure}

\begin{figure}[t]
  \centering
   \includegraphics[width=3.5in]{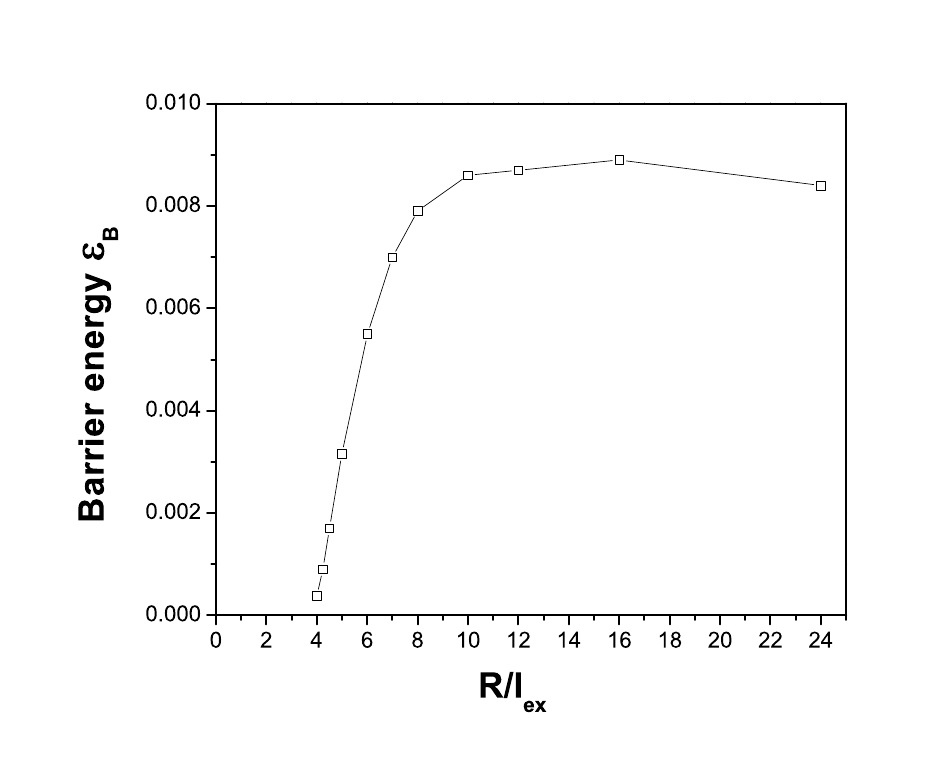}
\caption{ Barrier to vortex escape ($\epsilon_{B}$), defined by the relation
$ \epsilon_{B}=\epsilon_{max}-\epsilon_{vortex}$  as a function of dot
radius (in units of $l_{ex}$).}
   \label{fig:fig9}
\end{figure}

\begin{figure}[t]
  \centering
   \includegraphics[width=3.5in]{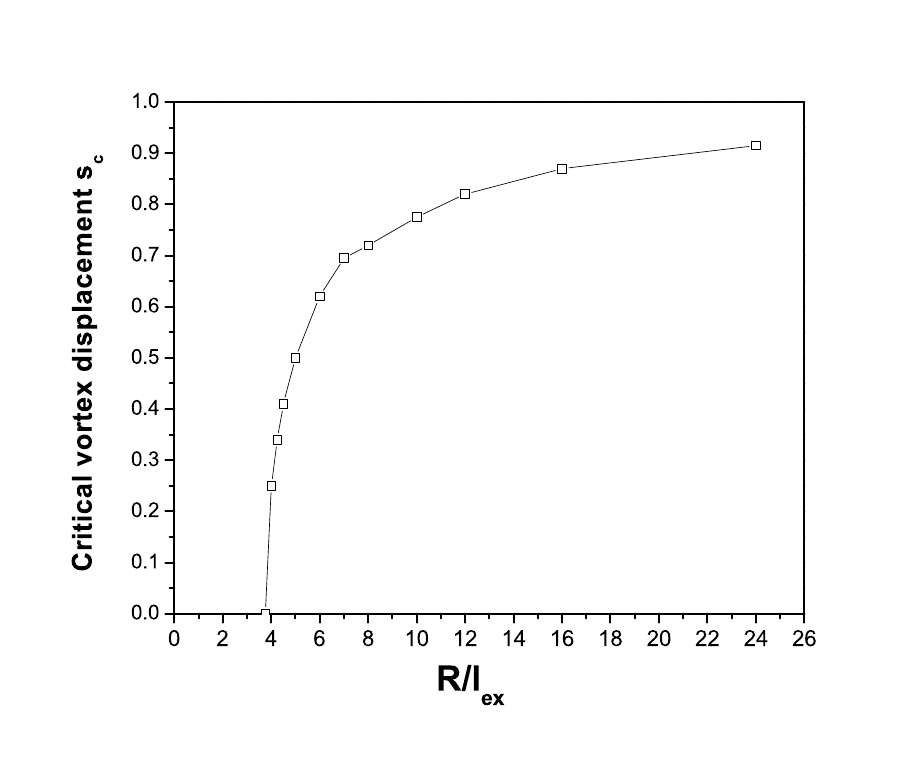}
\caption{Reduced displacement for maximum energy of the vortex state $s_{c}$ as a function of dot
radius (in units of $l_{ex}$).  }
   \label{fig:fig10}
\end{figure}

The maximum potential energy $E_{max}$ of the shifted magnetic vortex, evaluated from plots such as Fig.~\ref{fig:fig5}a, depends on the radius of the dot,
where it is confined. In Fig.~\ref{fig:fig8}, micromagnetic calculations of the reduced dot energy density $\epsilon$ are shown as function of dot radius $R$,
scaled by the exchange length. The curves correspond to the maximum vortex energy $E_{max}$ and
the minima associated with the three equilibrium states of the magnetization (in-plane,perpendicular,vortex).
 The vortex state is
unstable for small dots $R<R_{s}$, metastable for $R_{s}<R<R_{eq}$ and a ground state for $R>R_{eq}$ where
the values $R_{s}=2.5 l_{ex}$ and $R_{eq}=16 l_{ex}$ are obtained for the absolute and equilibrium single domain radius respectively. The corresponding variation of vortex barrier energy  $\epsilon_{B}=E_{max}-E_{vortex}$ and displacement
$l_{c}=s_{c}R$ for vortex escape are shown in
Figs.~\ref{fig:fig9} and \ref{fig:fig10} respectively. For small dots $R=3.5 l_{ex}$ , the vortex is  
unstable and any shift from equilibrium at the dot center 
results in relaxation
to the ground state (in-plane magnetization). The critical size for vortex instability is larger than
$R_{s}$ since the latter is defined assuming a random perturbation different than shifting the whole vortex. 
The vortex barrier energy increases for larger dots and attains 
a maximum value, for $R>10 l_{ex}$, related to the vortex annihilation field \cite{gus:01}.
The displacement $l_{c}$ for vortex escape  
 increases with disk size  to the maximum value imposed by the disk perimeter ($l_{c}/R \rightarrow 1$), 
attained for sub-micron dots $R>>l_{ex}$. For sufficiently large dots, the vortex is within the domain of attraction
of the dot center, irrespective of the initial position.  

\begin{figure}[t]
  \centering
   \includegraphics[width=3.5in]{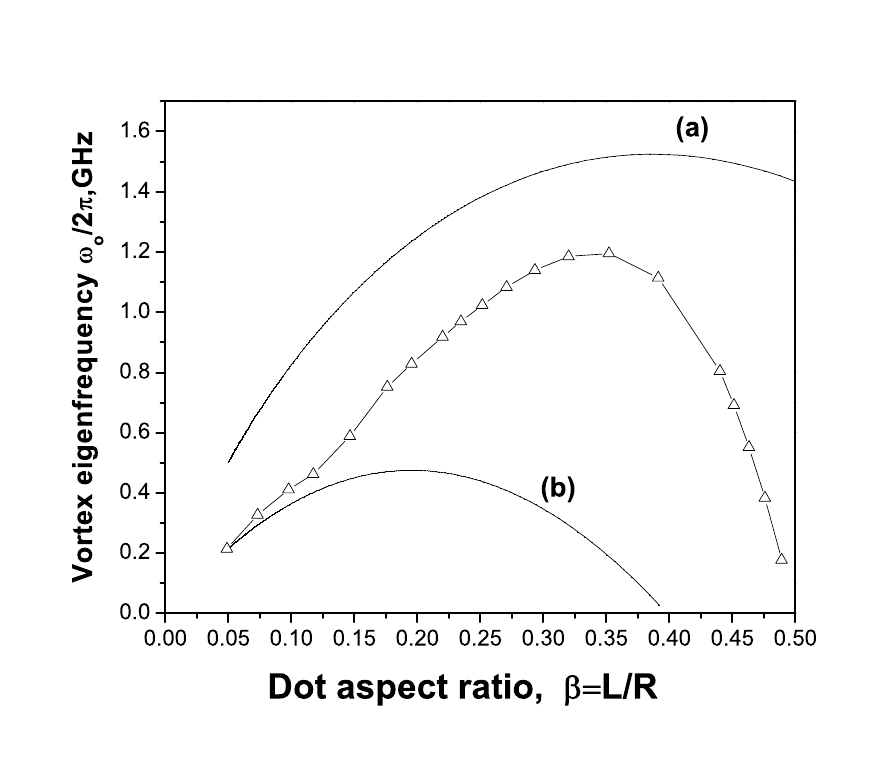}
\caption{Micromagnetic (markers) and analytical (solid lines) calculations of vortex  eigenfrequency $\omega_{o}/2\pi$
 as a function of dot aspect ratio $\beta=L/R$ a) the rigid vortex model and b) the two-vortices model. The
dot thickness is here $L=10$ nm.}
   \label{fig:fig11}
\end{figure}

 Micromagnetic calculations of the dependence of the fundamental vortex eigenfrequency $\omega_{o}/2\pi$, obtained for small perturbation $s<<1$ as in Fig.~\ref{fig:fig4}, on the
dot aspect ratio $\beta=L/R$  are shown in Fig.~\ref{fig:fig11}. The dot thickness was fixed ($L=10$ nm) and the radius $R$ was allowed
to vary. The eigenfrequency attains a maximum value at $\beta=0.35$ and vanishes
for smaller radius $\beta \simeq 0.5$ when the vortex becomes unstable. The maximum value arises from
the change in the relative contribution of the magnetostatic and exchange energy to the stiffness coefficient $\mu$.
For instance, the eigenfrequency assuming a 'rigid' vortex is \cite{gus:02}

\begin{equation}
 \omega_{o}=\frac{\pi\gamma M_{s}}{Q} \left[ F_{1}(\beta) - \frac{1}{(R/l_{ex})^{2} } \right]
\label{equ:o}
\end{equation}
where $F_{1}(x)=\int dt t^{-1} f(xt)J_{1}^{2}(t)$ corresponds to the averaged in-plane dot demagnetizing factor,
$f(x)=1-[1-\exp (-x)]/x$  and $J_{1}$ is the Bessel function. Using the approximation
$F_{1}(\beta) \simeq (\beta /2\pi) [\ln (8/\beta) ] -1/2 ]$, valid for $\beta<<1$, it can be shown that the eigenfrequency is maximum at radius $R=4\pi /L [\ln (8/\beta) -3/2 ]^{-1}$ where all lengths are in units of $l_{ex}$.
Previous studies were restricted to sub-micron sized dots where the second term of Eq.(\ref{equ:o}) could be neglected \cite{gus:02}, so a monotonically increasing eigenfrequency $\omega_{o}(\beta)$ arising from the magnetostatic
energy only was reported. 

The micromagnetic calculations are compared in Fig.~\ref{fig:fig11} with the curves obtained using the 'rigid' vortex
and 'two-vortices' approximations for the magnetostatic energy. For sub-micron sized disks with aspect ratio $\beta<0.05$ corresponding to a radius $R>200$ nm, the micromagnetic calculations are in good quantitative agreement with the 'two vortices' model, as reported in Ref.\cite{gus:02}.
In this regime, the eigenfrequency is determined primarily by the magnetostatic energy. The rigid vortex approximation fails to describe the dynamic behavior since the magnetostatic energy can be decreased by elimination of the surface 
charges at the disk perimeter at the expense of some contribution from volume magnetic charges arising from vortex deformation. For smaller disks $R<200$ nm, the vortex eigenfrequency is between the predictions of the two models. A reduction in side charges occurs but is not complete as a result of the large expense in exchange energy arising from  vortex deformation.  Similar results can be obtained in principle for thinner disks, however, the
 'two vortices' approximation is then valid for larger cylinders which are not easily amenable to micromagnetic simulations.

\section{Conclusions}

Micromagnetic calculations were carried out of the precessional behaviour of a single magnetic vortex or antivortex confined in a permalloy circular nanodot. The existence of two domains of attraction for the vortex state are identified
arising from a maximum in the potential energy of the shifted vortex. This effect is atrributed to the competition
between the magnetostatic attractive and exchange repulsive forces on the shifted vortex. Antivortices instead are always unstable and trace a spiral trajectory
of increasing distance from the dot center followed by annihilation at the dot envelope.
The precessional behaviour of vortices and antivortices is satisfactorily described by Thiele's theory
of collective coordinates, relating the angular frequency of precession to the gradient of the potential energy.
For small nano-sized dots, however, the 'rigid' vortex and 'two-vortices' approximation for the magnetostatic energy
is not satisfactory and the exchange forces have a significant effect on the translation mode vortex eigenfrequency. 
For antivortices, the development of an analytical model to account for the magnetostatic interaction in circular dots
is clearly needed to provide further insight on the results of our micromagnetic calculations.

Vortex stability is necessary in applications of nanostructured patterned media for data storage.
Microfabrication downscaling implies that the 
displacement $l_{c}$ and associated energy barrier may become useful characterization parameters
of the thermal stability of the recorded information.

% now the references. delete or change fake bibitem. delete next three
%   lines and directly read in your .bbl file if you use bibtex.
%\bibliographystyle{c:/latex/bibfiles/prsty}
%\bibliography{c:/latex/bibfiles/mainbib}
% figures follow here
%
% Here is an example of the general form of a figure:
% Fill in the caption in the braces of the \caption{} command. Put the label
% that you will use with \ref{} command in the braces of the \label{} command.
%

%\bibliographystyle{c:/latex/bibfiles/prsty}
%$\bibliography{c:/latex/bibfiles/mainbib}
% figures follow here
%
% Here is an example of the general form of a figure:
% Fill in the caption in the braces of the \caption{} command. Put the label
% that you will use with \ref{} command in the braces of the \label{} command.
%

\end{document}